\documentstyle{article}
\font\tenbf=cmbx10
\font\tenrm=cmr10
\font\tenit=cmti10
\font\elevenbf=cmbx10 scaled\magstep 1
\font\elevenrm=cmr10 scaled\magstep 1
\font\elevenit=cmti10 scaled\magstep 1

\renewenvironment{thebibliography}[1]
 { \elevenrm
   \begin{list}{\arabic{enumi}.}
    {\usecounter{enumi} \setlength{\parsep}{0pt}
     \setlength{\itemsep}{3pt} \settowidth{\labelwidth}{#1.}
     \sloppy
    }}{\end{list}}

\begin{document}
\noindent\hspace*{12cm} TPJU-33/94\newline
\hspace*{12cm} hep-ph/9412395  \newline
\begin{center}{{\tenbf
               HYPERINTERMITTENCY ? \footnote{ To be published in the
Proceedings
          of the XXIV International
       Symposium on Multiparticle Dynamics, Eds. A. Giovannini,
       S. Lupia and R. Ugoccioni, World Scientific, Singapore.  }  \\}
\vglue 5pt
\vglue 1.0cm
{\tenrm J. WOSIEK   \\}
\baselineskip=13pt
{\tenit Institute of Computer Science, Jagellonian University\\}
\baselineskip=12pt
{\tenit Reymonta 4, 30-059 Cracow 16, Poland\\}
\vglue 0.8cm
{\tenrm ABSTRACT}}
\end{center}
\vglue 0.3cm
{\rightskip=3pc
 \leftskip=3pc
 \tenrm\baselineskip=12pt
 \noindent
The analogy between intermittency and scaling in statistical physics
is extended to the case of more variables. It is shown that the
inclusive densities predicted by the perturbative QCD obey
generalized homogeneity principle which leads to many consequences in
statistical physics. In particular the hyperscaling
may be a prototype for the hyperintermittency in multiparticle
production at high energies.
\vglue 0.6cm}
{\elevenbf\noindent 1. Introduction}
\vglue 0.4cm
\baselineskip=14pt
\elevenrm
During the last few years the phenomenon of intermittency
has been intensively studied theoretically and experimentally.
These studies have originated from the attempt to  distinguish
dynamical and statistical fluctuations \cite{bp}. Simple models
with nontrivial dynamical fluctuations were proposed. They showed
characteristic power dependence of multiplicity moments on the volume
of the phase space. Therefore intermittency is often identified
with such a power behaviour. Also the terms self-similarity and
fractality were used in this context. It was later realized, that
the above power dependence is equivalent to the power behaviour of
the correlation functions in the corresponding range of variables
\cite{ja}.

It turned out that a good laboratory to study intermittency is provided
by statistical physics. In the latter intermittency is analogous
to scaling, or equivalently, to the existence of the nontrivial
anomalous dimensions of various field-theoretical operators. In
statistical systems this power behaviour occurs in the vicinity
of the  second order phase transition  and is specified by
a finite set of phenomenologically defined critical exponents.
Confirming the original motivation such a behaviour is driven by
the (thermal) fluctuations of arbitrary sizes. However in addition
to the simple scaling also the phenomenon of hyperscaling occurs
in statistical systems. This offers an interesting possibility to
extend the analogy between the intermittency in multiparticle
production and scaling in statistical systems to yet another level,
namely to attempt to define the counterpart of the hyperscaling
in multiparticle production. To this end we first review shortly
the idea of  the hyperscaling on the example of the two-dimensional
Ising model.

\vglue 0.6cm
{\elevenbf\noindent 2. Hyperscaling in statistical physics}
\vglue 0.4cm
 The vanishing of magnetization at the Curie point
defines the critical exponent $\beta$
\begin{equation}
m(t)\equiv <s> \sim t^{\beta},\;\;\;\; t=(T_c-T)/T_c > 0 . \label{mag}
\end{equation}
 Other exponents control divergence of the specific heat  \footnote{In the
particular case of the two-dimensional Ising model $\alpha=0$ and $c$
diverges logarithmically.}  $c$ ,
susceptibility $\chi$ and
of the correlation length $\xi$
\begin{eqnarray}
c(t) \sim |t|^{-\alpha}, \label{cv}  \\
\chi(t) \sim |t|{-\gamma} \\
\xi(t) \sim |t|^{-\nu}. \label{xi}
\end{eqnarray}
Connected correlation function also scales with the distance at
the critical point
\begin{equation}
\Gamma(r_1,r_2)\equiv <s_1 s_2>-<s>^2
\sim {1\over |r_1-r_2|^{d-2+\eta}}	 , \label{corr}
\end{equation}
where $d$ is the dimension of the system.
In general all observables depend on two variables: reduced temperature $t$
and external magnetic field $h$, e.g. $<s>=m(t,h)$.
All previous formulae apply for $h=0$.
When considered as the functions of the magnetic field, $h$,
magnetization also shows power behaviour
(at $T=T_c$) albeit with different exponent $\delta$.
\begin{equation}
m(0,h) \sim h^{1/\delta}, \label{mh}
\end{equation}
and silmilarly for other observables, Eqs. (\ref{cv}-\ref{xi}), but with
different
exponents.
Discussed above analogy with multiparticle production regards
power behaviour of the normalized factorial moments
\begin{equation}
M_q\equiv {<n^{(q)}>_{\Delta} \over <n>_{\Delta}^q} \sim \Delta^{-\alpha_q},
\end{equation}
as the counterpart of these scaling laws in statistical physics.
In this equation $<>_{\Delta}$ denotes averaging over the phase space region
of size $\Delta$.

It was observed \cite{wid} that the variety of scaling laws,
Eqs.(\ref{mag} - \ref{xi},\ref{mh} ) follow from the
single scaling law of the free energy density in two variables
\begin{equation}
f(\lambda^{y_t} t, \lambda^{y_h} h )=\lambda^d f(t,h), \label{gh}
\end{equation}
which is refered to as the generalized homogeneity principle \cite{di}.
Similar two-dimensional scaling was also postulated for the
correlation functions. The generalized homogeneity was later justified
theoretically using the renormalization group approach
\cite{kad,wil}.
 Note that there are only two independent exponents
$y_t,y_h$ in Eq.(\ref{gh}). All phenomenological exponents
(\ref{mag} - \ref{mh})
can be derived from these two and consequently there are relations between them
\cite{di}, \cite{thesis}.
There is an important class of relations between phenomenological exponents
of statistical systems which contains the dimensionality of the space $d$.
Such relations are called hyperscaling relations and their derivation requires
additional assumptions. The most widely known is the Josephson relation
which reads for the two-dimensional Ising model
\begin{equation}
\nu d = 2-\alpha. \label{jos}
\end{equation}
Hyperscaling, Eq.(\ref{jos}), has a simple interpretation proposed
by Pippard and Ginsberg . According to Eq.(\ref{cv}) the singular
part of the density of the free energy scales as
\begin{equation}
f(t) \sim |t|^{2-\alpha}, \label{free}
\end{equation}
which, using Eqs.(\ref{jos}) and (\ref{xi}) can be rewritten as
\begin{equation}
f(t)\sim {1 \over \xi(t)^{d}}. \label{xipower}
\end{equation}
 This means that the
change of the free energy due to the fluctuations of the size $\sim \xi$
is of the order 1 which may be intuitively expected \cite{di}.

Our aim is to explore a possibility of extending
the analogy between the scaling and the intermittency, and to attempt
to define the "hyperintermittency" in multiparticle spectra
as the counterpart of the hyperscaling in statistical systems.
In this lecture we shall restrict ourselves to the first step necessary in
such a construction, namely we will look for the analog of the generalized
homogeneity principle, Eq.(\ref{gh}), in the spectra of particles produced at
high energies.
\vglue 0.5cm
{\elevenbf \noindent 3. Generalized homogeneity in perturbative QCD}
\vglue 0.4cm
It turns out that the predictions of the perturbative QCD
for the intermittency in partonic cascade, have interesting properties
which are similar to the consequences of Eq.(\ref{gh}).
 Recently intermittency
in QCD jets was studied by  several groups \cite{we},\cite{dd},\cite{bmp}.
Together with W. Ochs we have calculated QCD predictions
 for the variety of partonic observables. They
reveal rather interesting universality and the new type of scaling behaviour
(see also the lecture by Wolfgang Ochs in these Proceedings).
\begin{eqnarray}
\rho(Q_1,Q_2) & \sim &  Q_1^{ a(Q_1) \omega(\epsilon) }, \label{qcd}
\\
\epsilon={\ln{Q_2} \over \ln{Q_1} }
\end{eqnarray}
where $\rho$ is the generic partonic density which depends essentially
on the two variables, (or scales) which are relevant to the particular process
 (see Table I for the list of studied reactions and for the definitions
of variables in various cases).    $a(q)$ is related to the coupling constant
$a(q)^2=6\alpha_s(q)/\pi$ and $\omega(\epsilon)$ denotes the calculable
scaling function which is largely process independent.

\begin{table}
\renewcommand{\arraystretch}{2}
\vspace*{3cm}
\begin{center}
\begin{tabular}{||c|c|c|c|c||} \hline\hline
 $\alpha_s$ & inclusive process  &  observable  &    $Q_1$  & $Q_2$  \\
  \hline\hline
 $constant$ & 1-parton   & $\rho^{(1)}_{(P,\Theta)}(k)$
     & ${P\Theta\over Q_0}$    & ${k\Theta\over Q_0}$           \\  \hline
 $running$ & 2-parton & $\rho^{(2)}_{(P,\Theta)}(\vartheta_{12})$
     & ${P\Theta\over\Lambda}$ & ${P\vartheta_{12}\over \Lambda}$ \\  \hline
 $running$ & n-parton & $h^{(n)}_P(\vartheta,\delta)$
     & ${P\vartheta\over\Lambda}$ & ${P\delta\over\Lambda}$  \\ \hline\hline
\end{tabular} \newline
\caption{Table I. Observables and corresponding kinematical variables
for the three processes studied: momentum distribution, distribution
of the relative angle, and the cumulant moment of n-th order respectively. }
\end{center}
\end{table}

It is readily seen that, similarly to the free energy, Eq.(\ref{gh}),
the generic density (\ref{qcd}) shows the  power behaviour
\footnote{Violated only by the running coupling constant.} along the special
family of curves
\begin{equation}
Q_2 = Q_1^{\epsilon}, \label{cur}
\end{equation}
in the plane $(Q_1,Q_2)$. In contrast to Eq.(\ref{gh}) the exponent
$\epsilon$ is not fixed ($ 0 < \epsilon < 1 $)  in QCD, while it is a single
number, $\epsilon=y_h/y_t$, in the Ising model.  This is the consequence of
more
complex behaviour of QCD under the renormalization group transformations.
Also, the simple
 $(\sim \lambda^d)$ scaling along the curve (\ref{cur}) in (\ref{gh})
 is replaced
by more complicated $\lambda^{a(Q_1)}$ one.
Nevertheless, the basic analogy between Eqs.(\ref{gh}) and (\ref{qcd})
is established.

To go futher and to define hyperintermittency in close
analogy to the hyperscaling relations, Eqs.(\ref{jos},\ref{xipower}), would
require more detailed analysis of the additional assumptions leading to
the latter in statistical models. This is beyond the scope of this lecture.
Since however the generalized homogeneity, Eq.(\ref{gh}), {\elevenit implies}
 a class of relations between critical exponents, one can forsee  that
similar class would follow from Eq.(\ref{qcd}) for partonic counterparts
of the magnetization, specific, heat, etc. Naturally, they follow only for
fixed $\epsilon$, therefore they will be less restrictive in the QCD case.
\vglue 0.6cm
{\elevenbf\noindent 4. Outlook}
\vglue 0.4cm

To conclude, we have shown that extending the intermittency study to
more variables, leads to more complete analogy with statistical
physics, which in turn could provide better understanding of multiparticle
production. QCD results \footnote{Derived in the double logarithmic
approximation.} exhibit generalized homogeneity well known in statistical
physics. Therefore the existence of the hyperintermittency seems plausible
and one may start searching for its implications.

It is interesting to point out one practical consequence of
hyperintermittency. The relation of type (\ref{jos}) allows
to determine the correlation length in the scaling regime.
Usually this parameter controls the exponential (or gaussian)
fall-off of the correlation functions. Near the critical point
however $\xi$ is very large and, consequently it is hard to
extract it from the very slow spacial decrease of the correlation
functions. Fortunately, hyperscaling provides a simpler way,
since, according to Eq.(\ref{xipower}), $\xi$ can be  directly
determined from the  {\elevenit temperature dependence} of some
observables.

\vglue 0.5cm
{\elevenbf \noindent 5. Acknowledgements \hfil}
\vglue 0.4cm
 We would like to mention the talk of Wu Yuanfang at this Symposium
\cite{liu} where the study of more complex scaling patterns in many variables
is proposed on more geometrical basis.

This work is partly supported by the KBN grants No PB 0488/P3/93 and
2P30Z2520G.
\vglue 0.5cm
{\elevenbf\noindent 6. References \hfil}
\vglue 0.4cm

\end{document}